\documentclass[12pt,twocolumn]{article}
\usepackage[a4paper,left=20mm,right=20mm,top=25mm,bottom=25mm,includeheadfoot]{geometry}

\usepackage{mathpazo} 
\usepackage{graphicx}
\usepackage{amsmath,textcomp}
\usepackage{makeidx}
\makeindex

\usepackage{sectsty}
	\allsectionsfont{\sffamily\raggedright}
	\sectionfont{\sffamily\large\raggedright}
	\subsectionfont{\sffamily\normalsize\raggedright}
    
\usepackage{cite} 

\usepackage[caption=false]{subfig}

\usepackage{widetext}
\usepackage{flushend}
\usepackage{cuted}
\usepackage{color}
\usepackage{graphicx}
\usepackage{hyperref} 
\usepackage{pstricks,amssymb}

\usepackage{lmodern} 
\usepackage{physics} 
\usepackage[output-decimal-marker={.}]{siunitx} 
\usepackage{systeme} 
\usepackage{url} 

\usepackage{fancyhdr}
\begin{document}
\title{\vspace{-2em}\bfseries\sffamily Estimating gravimetric effects for ordinary bodies}
\author{\normalsize Pier Franco Nali${^1}$ \\[2ex]
$^{1}$Independent scholar, Cagliari, Italy.\\
{\tt pfnali@alice.it}\\[2ex]
}

\date{\itshape Submitted on dd-MMM-yyyy}

\maketitle

\thispagestyle{fancy}

\begin{abstract}
{\sffamily
The main topic of this article is a discussion about the best way to show students that the proportionality of mass and weight, strictly true for point-like particles, is an excellent approximation for objects of “normal” size. The usual way of addressing this issue, although very simple, is not entirely satisfactory. Our approach considers first and second order, coordinate dependent, gravimetric effects, connected to the internal geometry of objects; these effects, extremely small, are estimated through examples.
}\\
\hrule
\end{abstract}

\section{Introduction} \label{sec:foo}
Direct proportionality of mass and weight is a well-established principle, proven as an experimental fact for all bodies in the same place. However, apart from the special case of uniform gravitational field, this principle is only valid locally, that is for point particles. When both the variability of the gravitational field and the bodies' internal structure cannot be ignored, the point-particle approximation fails and the proportionality of mass and weight cannot be regarded as strictly exact. Anyway, discrepancies involved are generally tiny and can safely be overlooked in most situations we commonly experience. 

It should be deemed that teachers usually show students that for common bodies the acceleration of gravity does not vary appreciably (or, otherwise said, the gravitational field is uniform) within the size of the object, which can be done very simply. At least at the college level, but also at the high-school level, after having presented relative motions and introduced the “apparent forces” in non-inertial reference frames - or “inertial forces” as we want to call them - teachers specify that the weight force on the earth's surface is the resultant of the (true) gravitational force and of the apparent (inertia) forces, in particular the centrifugal force if the body is stationary.

Limiting the problem to the gravitational component only (let's say it \(\vb{G}\)) teachers follow the usual simple path of differentiation \(\Delta G/G=\Delta R^{-2}/R^{-2}=-2\Delta R/R=-2h/R\), where \(R\) is the radius of the earth and \(h\) is the height of the body (or, if students do not know the differentials, simply calculate \(G(R+h)/G(R)=R^2/(R+h)^2\)). Since the difference in the first order is already very small, \(2h/R\) typically being of the order of \(10^{-8}\), that of the second order is obviously negligible for any practical effect. 

But we immediately realize that this method is flawed as the inverse proportion of the gravitational force to the square of the distance is strictly valid for point particles (and in the other special case of centrally-symmetric homogeneous bodies). In this way we implicitly assume the conclusion, as affirming uniformity of the gravitational field within point objects is tautological: our argument contains a circularity and runs into a logical fallacy (\textit{petitio principii}). Furthermore, properly speaking, \(h\) does not correlate to the object size: it merely represents a (small) displacement of the body (or rather, its center of mass) from the surface of the earth. 

Thus the argument is ineffective for extended bodies of arbitrary shape as it does not properly capture the effect on weight force (in either a rotating or a non-rotating frame) of the variation of \(\vb{g}\) within the object size; it would be preferable to find a different approach, allowing us to address the topic in broader generality and rigour.

\section{Background} \label{sec:bkg}
According to the “Declaration on the unit of mass and on the definition of weight; conventional value of \(g_n\)”: <<\textit{The word “weight” denotes a quantity of the same nature as a “force”: the weight of a body is the product of its mass and the acceleration due to gravity; in particular, the standard weight of a body is the product of its mass and the standard acceleration due to gravity.}>>\cite{CGPM} And, contextually: <<\textit{The kilogram is the unit of mass; it is equal to the mass of the international prototype of the kilogram.}>>\footnote{The definition of the kilogram in terms of the international prototype is obsolete and no longer in force since 20 May 2019; it has been redefined in terms of the Planck constant.} The value adopted for the standard acceleration due to gravity (on earth) is \(g_n=\SI{980.665}{\centi\meter\per\second\squared}\). 
Thus, for the weight force:
\begin{equation} \label{eq1}
\vb{w}=m\vb{g},
\end{equation}
where \(m\) is the mass of the object and \(\vb{g}\) is the acceleration vector due to gravity (We denote vectors, like \(\vb{g}\), \(\vb{w}\), \(\vb{G}\), as bold letters and represent their magnitudes, like \(g\), \(w\), \(G\), as italic letters).

More generally, weight means the gravitational force (or this plus the centrifugal force) on a small mass compared to that of the source (e.g. “the weight of astronauts on the Moon”). 

It is important to notice that “weight” and “gravitational force” are the same force but the use of either term is contextual and it is good practice to adhere to conventions on their use to avoid ambiguity. Calling the gravitational force on a celestial body “weight” creates confusion, and contradicts the convention that reserves this word for practical use (on weight vs gravitational force see e.g.\cite{Galili}). 

Hence the use of “weight” should be reserved for the force experienced by an object with mass in a gravitational field (e.g. light is bent by gravitational fields, although it has not weight, because it is massless).

We are now ready to introduce the alternate differentiation pathway  \(\Delta\vb{g}/g=1/g\left(\Delta\vb{g}/\Delta R\right)\Delta R=R/g\left(\Delta\vb{g}/\Delta R\right)h/R\), where \(h\) is now the linear size of the object and \(\vb{g}(R)\) (gravitational + centrifugal) does not have a form given \textit{a priori}. In this way we see that the less restrictive condition that the gradient \(\Delta\vb{g}/\Delta R\) is the same order of \(\vb{g}/R\) would fulfill the requirement of uniformity of the gravitational field within the size of the object. 

In a more formal way, if we think of the body made up of particles of masses \(m_i\), being \(m=\sum_{i} m_i\) the mass of the entire body and \(\rho(\vb{r})=\sum_{i} m_i\delta(\vb{r}-\vb{r}_i)\) its density, where \(\delta\) is the Dirac delta function, the weight force is the generalization of eq. \eqref{eq1} by integration over the whole space:
\begin{equation} \label{eq2}
\vb{w}=\iiint\rho(\vb{r})\vb{g}(\vb{r})\,\dd\tau=\sum_{i} m_i\vb{g}(\vb{r}_i).
\end{equation}
For an uniform field (\(\vb{g}(\vb{r})=g\vu{z}\)) the equation \eqref{eq2} reduces to \eqref{eq1} and the internal size and geometry of the body are irrelevant. 

But the uniformity condition for \(\vb{g}\) only holds approximately near the earth's surface. The earth's gravitational field is not uniform even on a small scale; modern gravimeters allow us to appreciate \(g\) with eight or nine significant digits (some \(\mu\)gals); this is how to say that variations in the earth's gravity between points even a few centimeters apart are detectable instrumentally. 

In this article we consider very small (some ppb) coordinate dependent effects. At these scales there are several others effects, both instrumental and environmental, which are not so weak. For example, \(\vb{g}\) has a dependence on time: the effect of terrestrial tides alone is two orders of magnitude greater (a few hundreds \(\mu\)gals), to which are to be added the effects of tides in the oceans, the hydrological and barometric components, also variable over time, and so on; also the motion (if any) of the measuring instrument has to be taken into account. Moreover, although independent of the mass \(m\) of the body under consideration, \(\vb{g}\) generally depends on “other” masses; it will be assumed that these external masses vary very slowly. 

Obviously we should not forget to mention the major non-gravitational contribution to weight, that of the centrifugal force due to diurnal rotation (which is a component of \(\vb{g}\)); the centrifugal force is some part per thousand of the gravitational force and the dependence of the two forces on the distance, from the center or from the axis, is different; the effect on the weight of Archimedes' thrust in the air is also significant. In the following we shall leave all these effects just mentioned aside from the present study and focus our attention on coordinate related ones, dependent on size and geometric configuration of objects. 

It is easy to see that for bodies for which experiments can be established, such as for bodies near the earth's surface, these coordinate-dependent effects are far too small for the standard resolution of dynamometers and scales. To this end we take Taylor series expansions truncated to the second-order of the functions \(\vb{g}(\vb{r}_ i)\) in eq. \eqref{eq2} centered about \(\vb{R}_{cm}=\frac{1}{m}\sum_i m_i \vb{r}_i\), that is the radius vector conducted from the origin of the coordinates to the body center of mass:\footnote{\label{note5}Under suitable analyticity conditions for the functions \(\vb{g}(\vb{r}_i)\).}

\begin{widetext}
\begin{equation} \label{eq3}
\vb{g}(\vb{r}_i)=\vb{g}_{cm}+\left(\vb{r}_i-\vb{R}_{cm}\right)\vdot\grad\vb{g}\bigr\rvert_{\vb{R}_{cm}}+\frac{1}{2}\left(\left(\vb{r}_i-\vb{R}_{cm}\right)\vdot\grad\right)^2\vb{g}\bigr\rvert_{\vb{R}_{cm}}
\end{equation}
(where \(\vb{g}_{cm}\) is evaluated in the center of mass).
\end{widetext}

By fixing the origin of the coordinates in the center of the earth, the equation \eqref{eq3} is quite exact for all applications in which we need to evaluate the weight force on bodies located on or near the earth's surface. In fact, under these conditions \(|\vb{r}_i-\vb{R}_{cm}|\ll R\) with \(|\vb{R}_{cm}|\cong R\) (assuming \(R\simeq\SI{6,371e6}{\meter}\) for the earth's mean radius) and the third and higher-order terms can be overlooked. 

\section{The first order effect. Implications for precision mass measurements} \label{sec:fnd}
For a (small) displacement of a body from \(P\) to a near point \(P^\prime\) we can express the variation of \(\vb{g}\) within the body by means of the eq. \eqref{eq3} as:
\begin{equation} \label{eq7}
\vb{g}(\vb{r^\prime}_i)=\vb{g(P^\prime)}+\left(\vb{r^\prime}_i-\vb{P^\prime}\right)\vdot\grad\vb{g}\bigr\rvert_{\vb{P^\prime}}
\end{equation}
(overlooking the small second order term). For a rigid body holds the distance preserving condition
\(\norm{\vb{r^\prime}_i-\vb{P^\prime}}\equiv\norm{\vb{r}_i-\vb{P}}\). For the sake of simplicity we assume the special condition of a purely translational displacement (preserving distance, angle, sense, and orientation) such that \(\vb{r^\prime}_i-\vb{P^\prime}\equiv\vb{r}_i-\vb{P}\), so that, if we take \(P\) as the center of mass, the second term of the right-hand side vanishes identically by introducing eq. \eqref{eq7} in eq. \eqref{eq2}, and eq. \eqref{eq7} reduces to
\begin{equation} \label{eq8}
\vb{g(P^\prime)}=\vb{g}_{cm}+\left(\vb{P^\prime}-\vb{R}_{cm}\right)\vdot\grad\vb{g}\bigr\rvert_{\vb{R}_{cm}}.
\end{equation}
In turn, eq. \eqref{eq2} reduces to
\begin{equation} \label{eq9}
\vb{w^\prime}=\vb{w}_{cm}+\vb{w}_{-1},
\end{equation}
where
\begin{equation} \label{eq10}
\vb{w}_{-1}=m\left(\vb{P^\prime}-\vb{R}_{cm}\right)\vdot\grad\vb{g}\bigr\rvert_{\vb{R}_{cm}}.
\end{equation}

Applying the gradient criterion \(\grad\vb{g}\sim \vb{g}/R\) we easily obtain \(w_{-1}/w_{cm}\sim h/R\), so, for a body similar in size and mass to the obsolete kilogram prototype (\(h\sim\) a few centimeters) 
\(w_{-1}\sim 10^{-7}-10^{-8}\) N.

In recent decades, in view of the redefinition of the SI units, in particular the kilogram, the goal set by the CGPM was to achieve accuracy of the order of \(10^{-8}\). Laboratories such as the National Physical Laboratory in the UK and numerous other metrology labs around the world have worked for years to achieve the required accuracy with Kibble's balances. At the meeting of 17\textsuperscript{th} May 2019 of the CCM, I. A. Robinson (NPL) stated: <<\textit{Whilst, at present, it is theoretically possible to measure the principal quantities to around 2--3 parts in \(\mathit{10^{9}}\) a number of other effects in the apparatus must be taken into account.}>>\cite{Robinson} This is a number of practical reasons, which limit accuracy, such as alignments, vibrations, etc. The NPL also plans to develop simpler Kibble's balances, affordable and operable in laboratories not as highly--specialized as NPL, capable of \(10^{-8}\) accuracy. At this accuracy level, a number of systematic effects has to be taken into account, including gravimetric contributions (see e.g. \cite {RobinsonSchlamminger}). An historical account of the development of these sensitive balances in the context of the proposed reform of the SI is outlined in \cite{Crease}. 

Apparently, gravimetric effects such as those we are talking about were first considered in the early 1970s in connection with the development at the National Bureau of Standards of the “One Kilogram Balance” NBS No. 2, whose standard deviation was approx. \(\SI{4}{\micro\gram}\).\cite{Almer}

Such kind of balances, used for comparing masses, compare the attractive gravitational forces between weights (or loads) and the earth. It is assumed (often implicitly) that these forces are exactly proportional to the masses of the loads (in vacuum) and do not vary during the measurement. The force on a standard weight used for the comparison of masses depends on the distance from the center of the earth to the center of gravity of the weight.\footnote {We anticipate here the notion of center of gravity that we will resume later. To practical effects of the discussion carried out in this section we can consider the center of gravity coincident with the center of mass, although the two concepts, in general, are to be kept distinct.} A second weight, of a different configuration, may have its center of gravity at a different distance from its base and thus the distance of the weight's center of gravity from the center of the earth will be different when the weight is placed on the weighing pan (which operates with the bases of the weights to be compared virtually on the same level). In this way, the constant of proportionality between the gravitational forces and the masses of the weights on the pan will be slightly altered, leading to a systematic error in the results of the comparisons between the masses, the so-called “gravitational configuration effect” introduced by Almer and Swift.\cite{AlmerSwift}  

If we consider a reference weight \(w_r=m_r g(R)\) and a second equal weight \(w_x\), whose centers of gravity are spaced by a distance \(\Delta h=d\) above their bases, then, from eqs. \eqref{eq9} and \eqref{eq10}:
\begin{align*} 
w_x=w_{x,cm}+w_{x,-1},
\end{align*}
\begin{align*} 
w_{x,cm}=m_x g(R),
\end{align*}
\begin{align*} 
w_{x,-1}=w_{x,cm}\cdot\frac{1}{g}\pdv{g}{h}\left(\Delta h\right),
\end{align*}
or, in the approximation of the gravitational component alone,
\begin{equation} \label{eq11}
w_{x,-1}=-w_{x,cm}\cdot\frac{2d}{R}.
\end{equation}
And, having imposed \(w_x=w_r\),
\begin{equation}\label{eq12}
m_x=m_r+m_{-1}=m_r+\frac{2d m_r}{R}.
\end{equation}

The term \(m_{-1}=2d m_r/R\) in eq. \eqref{eq12} is the (first order) corrective term that must be applied to the mass of the second weight to take into account the difference in the force of gravity on the weights placed on the weighing pan of the balance whose centers of gravity are at different distances from their bases. The corrective term can be evaluated independently of the equation \eqref{eq11}, valid in the approximation of the gravitational component alone, by directly measuring the acceleration of the free fall \(g\) and the gradient of the gravitational field \(\pdv*{ g}{h}\) in the place in which the mass calibration takes place. 

In the case \(d=\SI[retain-explicit-plus]{+1}{\centi\meter}\) the correction for the comparison of nominal weights of \(\SI{1}{\kilogram}\) calculated using the equation \eqref{eq12} is approx. \(\SI[retain-explicit-plus]{+ 3}{\micro\gram}\). The old Pt-Ir kilogram prototype (density \(\SI[per-mode=symbol]{21.55}{\kilogram\per\cubic\deci\meter}\)) is a right circular cylinder with a volume of approx. \(\SI{46.5}{\cubic\centi\meter}\) and approx. the same height (\(\SI{39}{\milli\meter}\)) as the diameter. Stainless steel samples (density \(\SI[per-mode=symbol]{8.00}{\kilogram\per\cubic\deci\meter}\)), having volume (\(\SI{125}{\cubic\centi\meter}\)) respecting the same proportions, have a height of \(\SI{54.2}{\milli\meter}\). The resulting distance of the samples' centers of mass (/gravity) from their base is higher than that of the prototype's center of mass from its base by an amount of \(\SI{7.6}{\milli\meter}\), which leads to a correction of \(\SI[retain-explicit-plus]{+2.4}{\micro\gram}\). For comparison, as Almer and Swift stated: <<\textit{Currently, mass comparisons at the 1-kg level can be carried out with standard deviations as small as 1.5 parts in 10\textsuperscript{9}.}>>\cite{AlmerSwift} 

This correction is far from being the most significant; the largest volume (\(\approx\SI{80}{\cubic\centi\meter}\)) of the stainless steel 1-kg samples results in a correction for the aerostatic thrust of approx. \(\SI[retain-explicit-plus]{+94}{\milli\gram}\) (assuming an air density of \(\SI[per-mode=symbol]{1.2}{\kilogram\per\cubic\meter}\)), that is about 40,000 times the gravitational effect.\cite{Jabbour} Nonetheless the gravitational correction becomes significant for high precision mass measurements. In fact, accuracy is limited not only by the achievable precision and uncertainty associated with the value of the sample, but also by systematic errors. It can be said that the accuracy of the results of the measurements is achieved only after all the relevant systematic errors have been identified and evaluated. This implies that in the design of an experiment all factors, even those that at first appear small, must be estimated to establish their potential importance as systematic factors affecting the measured results.
\section{The second order effect} \label{sec:soe}
By introducing eq. \eqref{eq3} into eq. \eqref{eq2} and noticing that the first order term vanishes identically for the choice of \(\vb{R}_{cm}\)\,, eq. \eqref{eq2} reduces to
\begin{align*} 
\vb{w}=\vb{w}_{cm}+\vb{w}_{-2},
\end{align*}
where
\begin{align*} 
\vb{w}_{cm}=m\vb{g}_{cm}
\end{align*}
and
\begin{align*} 
\vb{w}_{-2}=\frac{1}{2}\sum_i m_i\left(\left(\vb{r}_i-\vb{R}_{cm}\right)\vdot\grad\right)^2\vb{g}\bigr\rvert_{\vb{R}_{cm}}
\end{align*}
(or, in the continuous limit)
\begin{align*} 
=\frac{1}{2}\iiint \rho(\vb{r})\left(\left(\vb{r}-\vb{R_{cm}}\right)\vdot\grad\right)^2\vb{g}\bigr\rvert_{\vb{R_{cm}}}\,\dd\tau.
\end{align*}

The term \(\vb{w}_{cm}\) is the weight force acting on the material point to which the body is reduced, having the mass of the body and located in its center of mass. 

The term \(\vb {w}_{-2}\) is a second order gravimetric correction that takes into account the effect of the internal geometry of the body, estimated as follows:
\begin{align*} 
w_{-2}=w_{cm}\cdot\frac{1}{2}\cdot\frac{1}{g}\pdv[2]{g}{z}\left(\Delta z\right)^2,
\end{align*}
or, in the purely gravitational component approximation,
\begin{align*} 
w_{-2}=w_{cm}\cdot\frac{3d^2}{R^2}
\end{align*}
(where \(\Delta z=d\) is the linear size of the object). It represents the difference due to mass distribution around the center of mass compared to the situation in which all the mass is thought to be concentrated in one point. More formally, it can be shown (see e.g.\cite{Hestenes}) that the mass distribution intervenes to second order through the inertia tensor of the body. For a right circular cylinder of mass \SI{1}{\kilogram} a few centimeters high, like a copy of the old Pt-Ir kilogram prototype, the order of magnitude of the \(w_{-2}\) term is 
\(\sim 10^{-16}-10^{-17}\) N, the same of the weight of the equivalent mass of 1 joule, just 1⁄10 of that of the mass of an Escherichia coli bacterium and one hundred thousand times smaller than that of the mass of a human cell. 

Although fully negligible for bodies of ordinary mass near the surface of the earth, similar but a bit more significant effects occur in various kinds of problems, often faced with methods borrowed from celestial mechanics; in these situations, all the possible contributions must be carefully evaluated both in theoretical analyses and in the design of the experiments. A typical example are tidal phenomena, whose effects depend on the gradient of the gravitational field, rather than on intensity, and the variations of the gravitational force from one part of the object to the other must be considered. Meanwhile, there is no doubt that in these situations the bodies cannot be thought of as material points; Newton had already noticed that the exact results obtained for point-like particles are only approximate in presence of gravitational force between extended bodies attracting at short distances. In celestial mechanics it is usually satisfactory to stop calculations at the second order of approximation.

 Moreover, as the size of the objects under consideration are on a planetary or sub-planetary scale, i.e. a significant fraction (say, from a few thousandths to a few hundredths) of the earth's radius (think, for example, of lithosphere segments of which we want to study the isostatic conditions), or when the bodies are very close to an attracting center (a situation encountered in geophysical and astrophysical contexts), also the assumptions under which the equation \eqref{eq3} holds can fail and additional contributions should be considered. 
 
 In addition, sometimes it is not even possible to set up experiments or carry out direct measurements; when this occurs, the evaluation of gravitational forces needs \textit{ad hoc} modeling of objects, which may require, for example, the computation of quadruple or sextuple integrals and numerical integration (see, for example,\cite{Stirling}).
\section{The elusive center of gravity. Near--uniform field} \label{sec:fne}
The slight variation of the gravitational field within the size of earthly objects brings us to the interesting questions of the parallel field and the center of gravity.

The earth's gravitational field can be locally modeled by a field consisting of parallel vectors of (slightly) non-uniform intensity. This picture is useful because it allows us to introduce the “scalar weight” \(w\) in a coherent way,\footnote {“scalar” here does not mean invariant under rotation; here we intend 1-dimensional 1-component scalar field.} providing a tool to face and clarify the problem of determining a unique point (if any) where you can think applied the total weight force acting on all the particles of the body, i.e. its center of gravity.\footnote {The center of gravity is susceptible to other definitions, which we will not deal with here. A definition different from that of the weighted average can be given, for example, in the case of the spherically symmetric field.} A real gravitational field cannot be both parallel and non-uniform at the same time. It is convenient to examine the case of the near-uniform field, which, in addition to being simplistic, reproduces the gravitational field near the earth's surface with an excellent degree of approximation. Furthermore, with this choice, the problem can be dealt with in one dimension. For the usual central field
\begin{align*} 
\vb{g(r)}=-k\,\frac{\vb{r}}{\norm{r}^3}
\end{align*}
(\(k=GM\) for the earth's gravitational field) \(\div \vb{g}=0\) everywhere. In the near-uniform model we consider a small cylindrical region where there is a field of vectors parallel to \(\vu{z}\), having non-uniform modulus, so defined:
\begin{equation} \label{eq13}
\vb{g(r)}=g(z)\vu{z}=-kz^{-2}\vu{z},
\end{equation}
with \(z\gtrsim R\). 

The equation \eqref{eq13} does not represent a real Newtonian gravitational field as \(\vb{g}\) does not have zero divergence. However, for \(z\) large enough, i.e. far from the center of the field (e.g. near the earth's surface), the divergence is small and the eq. \eqref{eq13} is a very good approximation, locally (far from the center of the earth), of a gravitational field generated by a spherically symmetric mass distribution.\footnote {We assume the simplified picture of spherical earth, uniform density, not rotating; we abstract from all possible disturbing factors (assuming absence of air, no influence of celestial bodies, etc.). The ininfluence of the body under examination on the central gravitational field is also assumed (external field approximation).} In this framework, the center of gravity of a body can be defined through the “equipollent” moment condition (see \cite{Beatty}, p. 18). The moment of a single force on a particle is perpendicular to the force and the vector radius from the coordinate origin to the position of the particle. In general, however, this is not true for a system of forces; the total moment of a system of forces around a point \(O\) (the pole, which we will also assume as the origin of the coordinates) is generally not perpendicular to the total force vector acting on the system.

The moment \(\vb{T}_{eq}\) of the system of forces equipollent to a single weight force \(\vb{w}\) acting on the body satisfies the vector equation
\begin{equation} \label{eq14}
\vb{T}_{eq}=\vb{R}_{cg}\cross\vb{w},
\end{equation}
where \(\vb{w}\) is the total weight force acting on the body, defined by eq. \eqref{eq2} and \(\vb{R}_{cg}\) is the radius vector joining the pole with the point of application of this force, i.e. with the body center of gravity. The total moment of the forces acting on the system is by definition \(\vb{T}=\sum_i\vb{r}_i\cross\vb{w}_i\)\,, and the total weight force \(\vb{w}=\sum_i\vb{w}_i\)\,. Imposing the perpendicularity condition to these two vectors is equivalent to making the equation \eqref{eq14} valid for \(\vb{T}\), that we rewrite as
\begin{equation} \label{eq15}
\sum_i\left(\vb{r}_i-\vb{R}_{cg}\right)\cross\vb{w}_i=\vb{0}.
\end{equation}

The equation \eqref{eq15} (\textit {torque equation}) does not have solution if \(\vb{T}\) and \(\vb{w}\) are not orthogonal (and neither is zero) and in this case the center of gravity vector \(\vb{R}_{cg} \equiv (X, Y, Z)\) cannot be determined by this method. We do not examine here the existence conditions of the solutions of the \textit{torque equation}, whose detailed discussion can be found, for example, in \cite{Symon}. Fortunately, in the special case of parallel field the orthogonality condition is met.\footnote {Another case in which this condition is met is that of a planar system of forces.} If we choose the \(z\)--axis in the direction of the field, then \(\vb{w}_i=w_i\,\vu{z}\) and eq. \eqref{eq15} reduces to the linear system 
\[
\left\{
\begin{array}{lrcrcr}
\sum_{i} & (x_{i} & - & X)\quad w_{i} & = & 0\quad,\\
\sum_{i} & (y_{i} & - & Y)\quad w_{i} & = & 0\quad.
\end{array}\right. 
\]

The moment of total weight force will have only the \(x\) and \(y\) components different from zero, from which the \(X\) and \(Y\) components of the vector \(\vb{R}_{cg}\) can be calculated; these define the line of action of the total weight force. There remains the \(z\) component to be determined (the \textit{torque equation} for the \(z\) component is a null identity). We observe, however, that under the assumptions made the equation \eqref{eq15} can be rewritten as
\begin{equation} \label{eq16}
\left(\sum_i w_i\vb{r}_i-w\vb{R}_{cg}\right)\cross\vu{z}=\vb{0}.
\end{equation}

Then, as the pole \(O\) can be chosen arbitrarily and \(\vu{z}\) is a fixed vector, the equation \eqref{eq16} can be satisfied by choosing the vector \(\vb{R}_{cg}\) defined as (see \cite {Beatty}, p. 48)
\begin{equation} \label{eq17}
\vb{R}_{cg}=\frac{1}{w}\sum_i w_i\vb{r}_i=\frac{1}{w}\sum_i m_i g\left(\vb{r}_i\right)\vb{r}_i
\end{equation}
or, in the continuous limit,
\begin{equation} \label{eq18}
\vb{R}_{cg}=\frac{1}{w}\iiint \rho(\vb{r})g\left(\vb{r}\right)\vb{r} \,\dd\tau,
\end{equation}
which constitute the definition of the center of gravity in the case of parallel field. For a uniform field the equation \eqref{eq18} becomes
\begin{equation} \label{eq19}
\vb{R}_{cm}=\frac{1}{m}\iiint \rho(\vb{r})\vb{r} \,\dd\tau,
\end{equation}
that is \(\vb{R}_{cg}\) coincides with the center of mass radius vector. In the equations \eqref{eq18} and \eqref{eq19} it is implied that
\begin{align*}
w&=\iiint \rho(\vb{r})g(\vb{r}) \,\dd\tau,\\ m&=\iiint \rho(\vb{r}) \,\dd\tau.  
\end{align*}

With series expansions of \(w\) and \(g\left(\vb{r}\right)\) in eq. \eqref{eq18} around the center of mass,\footnote{See previous note \ref{note5}.} using the equations \eqref{eq3} and \eqref{eq13}, and truncating after the first order, we have (we omit the detailed steps):

\begin{align*}
\vb{R}_{cg}=\vb{R}_{cm}+\vb{R}_{-1}+\cdots,
\end{align*}

\begin{widetext}
\begin{equation} \label{eq20}
\vb{R}_{-1}=-\frac{2}{m Z_{cm}}\iiint \rho\left(\vb{r}\right)\left(z-\vb{R}_{cm}\vdot\vu{z}\right)\left(\vb{r-R}_{cm}\right) \,\dd\tau.
\end{equation}
\end{widetext}
If as an example we consider a solid in the shape of a right cylinder or a rectangle parallelepiped, very elongated with respect to its basis, resting on the earth's surface so as to approach the situation of a parallel and near-uniform field, we reduce the problem to one dimension. If \(h\) is the height of the solid, the \(z\)--coordinate of its center of mass will be given by \(Z_{cm}=R+h/2\); we also express the variable of integration as a function of the coordinate in the system of the center of mass \(\zeta=z-Z_{cm}\); finally, for simplicity, suppose the solid of uniform density \(\rho\). Then we can write the \(Z\) coordinate of the center of gravity as 

\begin{widetext}
\begin{align*}
Z=Z_{cm}+Z_{-1}+\cdots=Z_{cm}-\frac{2}{hZ_{cm}}\int_{-h/2}^{h/2}\zeta^2\,\dd\zeta+\cdots=Z_{cm}-\frac{h^2}{6Z_{cm}}+\cdots.
\end{align*}
\end{widetext}

The term \(Z_{-1}=-h^2/6Z_{cm}\cong-h^2/6R\) in the first order of approximation represents the displacement of the center of gravity apart from the center of mass. This is a tiny difference: in the case of Dubai's Burj Khalifa, currently the tallest building in the world (\(h=\SI{829.80}{\meter}\)), the center of gravity is only about \(\SI{2}{\centi\meter}\) below the center of mass! The center of gravity is a specially elusive concept. It identifies a defined point, but, unlike the center of mass, it does not have a definite position. Its position depends, in general, on the relative positions of the body under consideration and the attractive mass. As can be seen from the equation \eqref{eq20}, when the distance \(Z_{cm}\cong R\) of the body from the center of the earth increases, the center of gravity approaches the center of mass. This feature makes it difficult to work with the center of gravity and in practice this concept is seldom used. The detailed treatment of this and other interesting problems related to the center of gravity is beyond our scope; an introductory discussion on these topics can be found on the Wikipedia page \textit{“Centers of gravity in non-uniform fields”}\footnote{\url{https://en.wikipedia.org/wiki/Centers_of_gravity_in_non-uniform_fields}} and related talk, \footnote{\url{https://en.wikipedia.org/wiki/Talk:Centers_of_gravity_in_non-uniform_fields}} to which the interested reader is referred. 
\section{Conclusions} \label{sec:fnf}
We have established that the proportionality of mass and weight for ordinary bodies can be taken as an excellent approximation in all cases of practical interest. However, it is advisable for students to always clarify the limits of validity of this approximation, both in their theoretical meaning and for the aspects related to the sensitivity of the experiments. For this purpose, the gradient criterion \(\Delta\vb{g}/\Delta R\sim\vb{g}/R\) is suitable for exploring the variation of the gravitational force within the size of the body. While it is easy to show that this gravimetric effect is negligible for ordinary bodies, special caution should be observed when, in investigating certain areas, you go beyond the validity range of the point particle approximation. In geophysics, hydrostatics and astrophysics various situations are encountered of strongly inhomogeneous gravitational field and the gravitational effects connected to the internal geometry of the bodies cannot be neglected. Such effects must be carefully considered; for example: in celestial mechanics and astrodynamics, in the calculation of the short-distance interaction of non-spherical shaped bodies (see, e.g., \cite{Ashenberg, Shi, Hou}); in geophysics, in the calculation of the gravimetric field of a polyhedral plate (see, e.g., \cite{Banerjee, Nagy, Karcol}); in hydrostatics, in the computation of the thrust, where the pressure gradient is replaced by the product of the density of the fluid and the gravitational field (see, e.g., \cite {Lima}). These problems are addressed on a case-by-case basis and often require the development of specific solutions.

\section*{Acknowledgments}
I thank Santo Armenìa for drawing my attention to this topic. I also thank the reviewer for helpful comments/suggestions.

\end{document}